\documentclass[aps,prl,twocolumn,floats,showpacs]{revtex4}
\usepackage{epsfig}

\begin{document}

\title{Distribution of the local density of states,
reflection coefficient and Wigner delay time in absorbing ergodic systems
at the point of chiral symmetry}
\vskip 0.5cm
\author{Yan V. Fyodorov$^{1,*}$ and Alexander Ossipov$^{2}$}

\address{$^1$Department of Mathematical Sciences, Brunel
University, Uxbridge UB83PH, United Kingdom}
\address{$^2$ Max-Plank-Institut f\"{u}r Str\"{o}mungsforschung, 37073
G\"{o}ttingen, Germany}

\date{\today}

\begin{abstract}
Employing the chiral Unitary Ensemble of random matrices
we calculate the probability distribution of the local
density of states for zero-dimensional
("quantum chaotic") two-sublattice
systems at the point of chiral symmetry $E=0$ and in the presence of
uniform absorption. The obtained result can be used to
find the distributions of the reflection coefficent and of the
Wigner time delay for such systems.

\end{abstract}

\maketitle

Spectral and transport properties of quantum systems with chiral symmetries
attracted a considerable attention recently\cite{Br,Tit,AM,off,rflux,QCD,TI,EK}.
The systems of the discussed type are characterized by
random Hamiltonians of the form:
\begin{equation}\label{hform}
{\cal H}_{ch}=
\left(\begin{array}{cc} 0 & \hat{H}\\
\hat{H}^{\dagger} & 0 \end{array}\right)
\end{equation}
serving to describe various physical situations, e.g.
lattice systems with bond disorder of pure nearest-neighbour
hopping type\cite{Br,Tit,AM,off}, random magnetic flux models
appearing in the context of the Quantum Hall effect\cite{rflux}, as
well as lattice QCD models\cite{QCD}. Many other interesting examples can be
found in the cited literature.

Replacing the block $\hat{H}$ in Eq.(\ref{hform}) with random rectangular
matrices of the size $N_A\times N_B$ with independent,
Gaussian-distributed complex entries we arrive at
the object known as the chiral
Gaussian Unitary Ensemble (chGUE)\cite{AZ}.
In the language of the two-sublattice disordered lattice system
this model is known to correspond to the so-called universal ergodic
(or "zero-dimensional", or "quantum dot") limit adequate if we are interested in energies
smaller than the so-called Thouless energy $E_c$. The latter
requirement means that the time to diffusively propagate through the
system is much shorter than $\hbar/\Delta$, with $\Delta$
standing for the mean level spacing in the relevant energy range.
Spectral statistics of chGUE matrices is well-studied by various methods
and many explicit results are available\cite{ch1,ch2,ch3,iv,ch4,ch5}.

One of the most striking features of the spectra of chiral systems
is a specific behaviour of the spectral characteristics at zero
energy. In zero-dimensional chiral systems this specific behaviour
is manifested, in particular, via vanishing of the mean eigenvalue
density $\rho_{av}(E)$ at the middle of the band, $E=0$, as long as $N_A=N_B$.
For $N_A\ne N_B$ there exist exactly $|N_A-N_B|$ zero eigenvalues.
All nonzero eigenvalues for chiral models appear
in pairs $\pm \lambda_n$, with the
corresponding eigenvectors being $| w_n^{\pm}\rangle=\left(\begin{array}{c}
{\bf u}_n\\ \pm {\bf v}_n\end{array}
\right)$. Here ${\bf u}_n$ and ${\bf v}_n$ are eigenvectors of the
Hermitian matrices $\hat{H}\hat{H}^{\dagger}$ and
$\hat{H}^{\dagger}\hat{H}$, of the size $N_A\times N_A$ and
$N_B\times N_B$,  respectively. To ensure the
orthonormality of the eigenvectors
$\left\langle w_m^{\pm}|w_n^{\pm}\right\rangle=\delta_{mn}$ we choose the
normalisation
${\bf v}^{\dagger}_n{\bf v}_n={\bf u}^{\dagger}_n{\bf u}_n=1/\sqrt{2}$.

We first assume, for simplicity, that the number of lattice sites in the two
sublattices are equal: $N_A=N_B=N$ and define in a usual way the exact {\it
local} density of states (LDOS) at a site $j_A=1,2,...,N$ of the sublattice
$A$ via the relation:
\begin{eqnarray}\label{rhoa}
\rho_{j_A}(E_{\eta})=\frac{1}{\pi}\mbox{Im}
\langle j_A|\frac{1}{E_{\eta}-{\cal H}}|j_A\rangle \\
=\frac{2}{\pi}\mbox{Im}\sum_{n=1}^N |u_n(j)|^2 \frac{E_{\eta}}
{E^2_{\eta}-\lambda^2_n}
\end{eqnarray}
where $u_n(j)$ stands for $j-th$ component of the eigenvector
${\bf u}_n$. Here we assume that the energy $E_{\eta}=E-i\eta$ is
complex, the parameter $\eta>0$ standing for the level broadening
due to an absorption, i.e. uniform-in-space losses of the flux of particles
in the sample. In particular, for the energy $E$ chosen precisely in the
center of the spectrum ("the point of chiral symmetry") the LDOS
is given by the most simple expression:
\begin{equation}\label{start}
\rho_{j,A}(\eta)
=\frac{2}{\pi}\sum_{n=1}^N |u_n(j)|^2 \frac{\eta}
{\eta^2+\lambda^2_n}.
\end{equation}
For the lattice sites of the sublattice $B$ the LDOS
is given by the same formula, with ${\bf u}_n(j)$ replaced by
${\bf v}_n(j)$.

The above expression is a very convenient starting point for calculating the
distribution function of the local DOS at the spectrum center.
In doing this we employ the
method suggested for GUE without chiral structure in\cite{GUE}.
Namely, we consider the generating function (the Laplace transform of
the probability density) $P(t)=\left\langle
e^{-t \rho_{j,A}}\right\rangle_{chGUE}$ where the angular
brackets stand for the ensemble averaging. Using that the
eigenvalues $\lambda_n$ are statistically independent of the
eigenvectors ${\bf u}_n,{\bf v}_n$, and the components of
the latter vectors behave in
the large $N$ limit like independent complex Gaussian variables
with zero mean and variances $\left\langle
|u_n(j)|^2 \right\rangle_{chGUE}=\frac{1}{2N}$ one can first
perform the ensemble average over eigenvectors.
After simple manipulations one arrives at the following representation
for the generating function:
\begin{equation}\label{det}
P(t)=\left\langle
\frac{\det{\left(\eta^2+\hat{H}\hat{H}^{\dagger}\right)}}
{\det{\left(\tilde{\eta}^2+\hat{H}\hat{H}^{\dagger}\right)}} \right\rangle_{H}
\end{equation}
where we denoted $\tilde{\eta}=\sqrt{\eta^2+\eta t/(\pi N)}$.
In this way the problem of calculating the generating function in
question is reduced to
evaluating the ensemble average of the ratio of characteristic
polynomials of the random matrices $\hat{H}\hat{H}^{\dagger}$.
The latter object (and its generalisations) were intensively studied
in the literature recently, see \cite{ch4,ch5}, and the result
is readily available:
\begin{eqnarray}\label{avres}
P(t)=\frac{\pi}{2}\left[\epsilon I_1(\epsilon)K_0
\left(\sqrt{\epsilon(\epsilon+2 t/\pi)}\right)\right. \\ \nonumber
\left.+\left(\sqrt{\epsilon(\epsilon+2 t/\pi)}\right)
K_1\left(\sqrt{\epsilon(\epsilon+2 t/\pi)}\right)I_0(\epsilon)\right]
\end{eqnarray}
where we denoted $\epsilon=2N\eta$ and considered
this parameter to be finite when performing the large-N limit.
Here $I_n(z),K_n(z)$ stand
for the modified Bessel and Macdonald functions
of the order $n$, respectively. The distribution function
${\cal P}(\rho_A)$ of the LDOS
can be found by inverting the Laplace transform Eq.(\ref{avres}) with respect to $t$
and is given by:
\begin{equation}\label{main1}
{\cal P}(\rho_A)=
\frac{\epsilon}{2\rho_A}e^{-\frac{\epsilon}{2}(\pi\rho_A+1/\pi\rho_A)}
\left[I_1(\epsilon)+\frac{1}{\pi\rho_A}I_0(\epsilon)\right]
\end{equation}

Having the LDOS distribution at our disposal, we can further utilise
it for obtaining distributions of varous physical quantities of much
interest describing scattering of a quantum particle from the
two-sublattice disordered sample. In doing this we follow the method
suggested recently by the first author in \cite{my1}. Consider a particle
injected into the site $j_A$ of the sublattice $A$ through a
one-channel infinite lead perfectly coupled to the system.
The scattering of such a particle from
our absorptive sample is described by the scattering
matrix $S_{jA}$ expressed in terms of the matrix elements of the resolvent
associated with the chiral Hamiltonian ${\cal H}$:
\begin{equation}\label{smatrix}
S_{jA}(E_\eta)=\frac{1-iG}{1+iG}\quad,\quad G=\pi \langle j_A|
\frac{1}{E_\eta-{\cal H}}|j_A\rangle.
\end{equation}
In this way the problem of investigating statistics of the
scattering matrix amounts to that
of the diagonal entry $G$ of the resolvent. For an arbitrary energy $E$
the variable $G$ is
a complex quantity and one has therefore to know the joint
probability density of its real and imaginary parts.
However, at the point $E=0$ of chiral symmetry $G$ turns
out to be purely imaginary and therefore just proportional to the
LDOS, Eq.(\ref{start}), whose distribution ${\cal P}(\rho_A)$ was calculated
above, Eq.(\ref{main1}).
As a result we can calculate the distribution of the reflection
coefficient $R_{jA}(\eta)=|S_{jA}(\eta)|^2$. Below we present the
distribution of a related object $\tau=1-R_{jA}=4\pi\rho_{jA}/(1+\pi\rho_{jA})^2$
which, following
\cite{my1}, we suggest to call the "probability of no return" (PNR).
This quantity has meaning of the quantum-mechanical probability for a
particle entering the system via a given channel never exit
back to the same channel. It is well-defined for a given
realization of disorder and will therefore show sample-to-sample
fluctuations. Let us stress that
in the case $\eta=0$ of no internal dissipation $S-$matrix is unitary
for any energy and PNR is vanishing identically.
For a system with absorption the unitarity is violated,
giving rise to a nontrivial statistics of $\tau$:
\begin{equation}\label{main2}
{\cal P}(\tau)=
\frac{\epsilon}{\tau\sqrt{1-\tau}}e^{-\epsilon\left(\frac{2}{\tau}-1\right)}
\left[I_1(\epsilon)+\left(\frac{2}{\tau}-1\right)I_0(\epsilon)\right]
\end{equation}

\begin{figure}[t]
\epsfig{figure=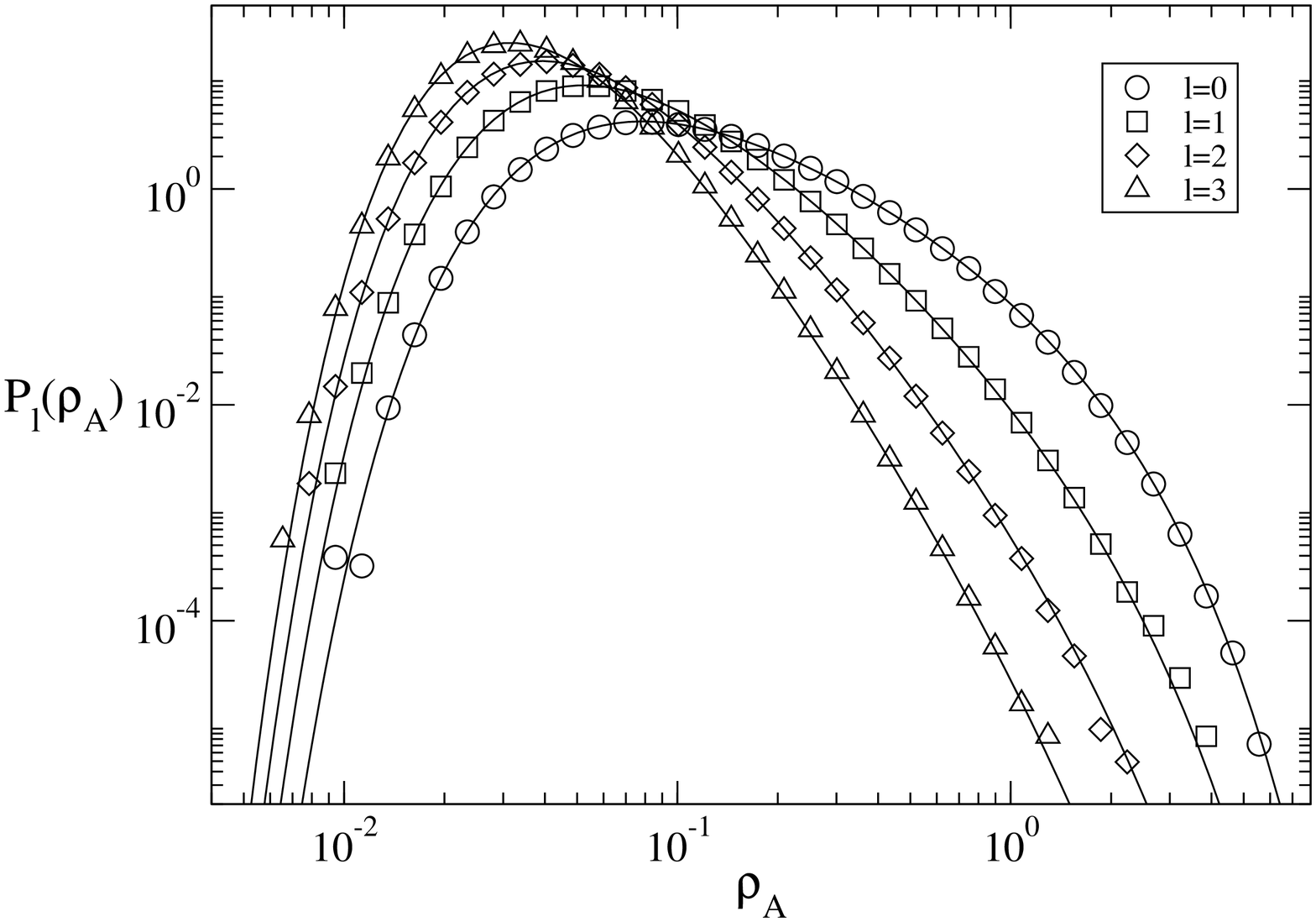,clip=,scale=0.33,angle=0}
\caption{\label{fig-ldos_A} Numerical (symbols) and analytical
(solid lines) results for the LDOS distribution of the sublattice
$A$ for $\epsilon =1$, $N_B=300$ and $N_A=300-l,\; l=0,1,2,3$.}
\end{figure}

The above distribution can be further utilized for finding
statistics of such interesting characteristic of quantum
scattering as the so-called Wigner delay time $\tau_W=-i\partial
S/\partial E$ intensively studied in recent years, see
\cite{td1},\cite{td2},\cite{tsamp},\cite{td3} and references
therein. This can be done again following \cite{my1}, see also
\cite{td3}. Indeed, remembering that nonzero values of $\tau$
arise solely due to an absorption, one can relate for small
absorption the PNR value to the Wigner delay time by $\tau\simeq
2\epsilon \tau_{W}$\cite{td2}. Taking the corresponding limit in
the distribution (\ref{main2}) we arrive at the distribution of
the scaled variable $\tilde{\tau}_W=\tau_W/2N$:
\begin{equation}\label{main3}
{\cal P}(\tilde{\tau}_W)=
\frac{1}{\tilde{\tau}^2_W}e^{-\frac{1}{\tilde{\tau}_W}}
\end{equation}
The expression Eq.(\ref{main3})) should be compared with the
corresponding formula
 for the one perfect channel attached to an ergodic system modelled by
 GUE, i.e. without the
chiral symmetry \cite{td1}: ${\cal P}(\tilde{\tau}_W)=\frac{1}{
\tilde{\tau}^3_W}e^{-\frac{1}{\tilde{\tau}_W}}$. The crossover
between the two types of behavior is expected to happen in the
present model with increasing energy parameter $E$ from zero to
values much exceeding the mean level spacing. Investigating this
interesting question requires however much more elaborate methods
and is left for future research.

\begin{figure}[t]
\epsfig{figure=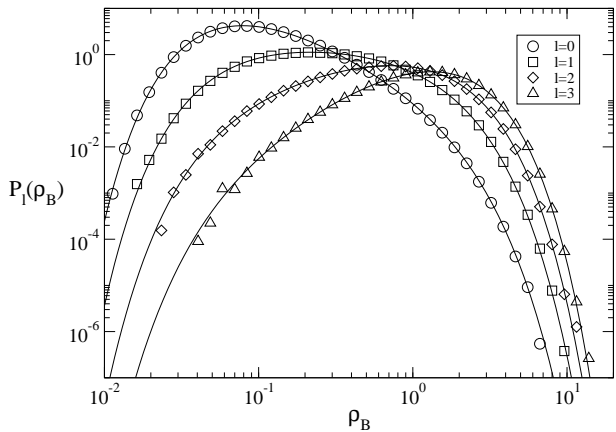,clip=,scale=0.33,angle=270}
\caption{\label{fig-ldos_B} Numerical (symbols) and analytical
(solid lines) results for the LDOS distribution of the sublattice
$B$ for $\epsilon =1$, $N_B=300$ and $N_A=300-l,\;
l=0,1,2,3$.}
\end{figure}

 Let us now briefly discuss generalization of our results to the case
of non-equivalent sublattices $N_B>N_A$. Note that in this case
exactly $N_B-N_A=l$ zero modes are supported by the sublattice $B$,
whereas only modes with nonzero eigenvalues give contribution to the LDOS
for the sublattice $A$. Assuming that $l\ll N_A,N_B$
one finds that LDOS distribution for the sites belonging to
the sublattice $A$ is given by
formula generalizing Eq.(\ref{main1}):
\begin{equation}\label{mainl}
{\cal P}_l(\rho_A)=
\frac{\epsilon}{2\pi^{l}\rho_A^{l+1}}
e^{-\frac{\epsilon}{2}(\pi\rho_A+1/\pi\rho_A)}
\left[I_{l+1}(\epsilon)+\frac{1}{\pi\rho_A}I_l(\epsilon)\right]
\end{equation}
whereas the LDOS distribution ${\cal P}_l(\rho_B)$ for sites
belonging to the sublattice $B$ is obtained from the above formula by
replacing $l$ with $-l$. In Fig.~(\ref{fig-ldos_A}) and Fig.~(\ref{fig-ldos_B})
we compare this prediction with the results of direct numerical simulations
of the random matrix Hamiltonian Eq.(\ref{hform}), for
sublattices $A$ and $B$ correspondingly. It is instructive to compare
 the behavior of the mean
LDOS for the two sublattices in the  limit $\epsilon\to 0$:
\begin{eqnarray}\label{meanl}
\langle\rho_A\rangle=
\frac{\epsilon}{2}
\left[I_{l+1}(\epsilon)K_{l-1}(\epsilon)+
I_l(\epsilon)K_{l}(\epsilon)\right]\to \frac{\epsilon}{4l}\\
\langle\rho_B\rangle=\frac{\epsilon}{2}
\left[I_{l-1}(\epsilon)K_{l+1}(\epsilon)+
I_l(\epsilon)K_{l}(\epsilon)\right]\to \frac{l}{\epsilon}
\end{eqnarray}
The singularity in the latter expression
 reflects presence of exactly $l$ zero
modes supported by the sites of sublattice $B$.

\begin{figure}[t]
\epsfig{figure=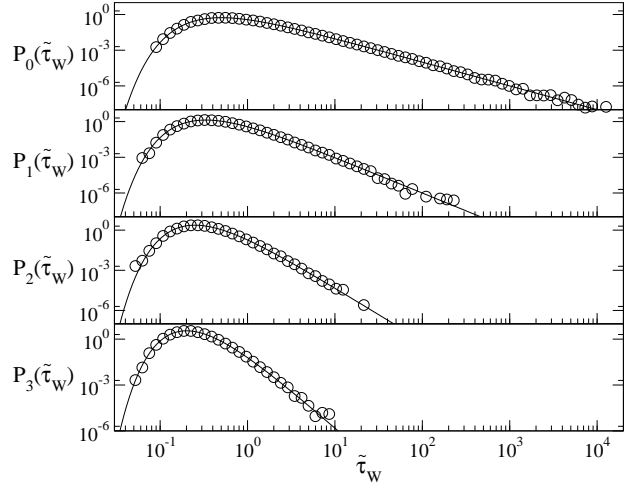,clip=,scale=0.33,angle=270}
\caption{\label{fig-tau} Numerical ($\circ$) and analytical (solid
lines) results for the distribution of scaled Wigner delay times
for sublattice $A$, for $\epsilon =1$, $N_B=300$ and $N_A=300-l,\;
l=0,1,2,3$.}
\end{figure}

After straightforward manipulations one can find that the distribution
Eq.(\ref{main2}) is replaced by the following two formulas:
\begin{eqnarray}\label{main2l}
{\cal P}_l(\tau_A)=
\frac{\epsilon}{\tau_A\sqrt{1-\tau_A}}
e^{-\epsilon\left(\frac{2}{\tau_A}-1\right)}
\\ \nonumber \times
\left[\left(v_{+}^l+v_{-}^l\right)I_{l+1}(\epsilon)+
\left(v_{+}^{l+1}+v_{-}^{l+1}\right)
I_l(\epsilon)\right],\\
{\cal P}_l(\tau_B)=
\frac{\epsilon}{\tau_B\sqrt{1-\tau_B}}
e^{-\epsilon\left(\frac{2}{\tau_B}-1\right)}
\\ \nonumber \times
\left[\left(v_{+}^l+v_{-}^l\right)I_{l-1}(\epsilon)+
\left(v_{+}^{l-1}+v_{-}^{l-1}\right)
I_l(\epsilon)\right].
\end{eqnarray}
valid for the sublattices $A$ and $B$, respectively.
Here the parameters $v_+,v_-$ satisfy the conditions: $v_+v_-=1$
and $\frac{1}{2}(v_{+}+v_{-})=\left(\frac{2}{\tau_{A(B)}}-1\right)$.
Finally, performing in Eq.(\ref{main2l}) the limiting procedure
discussed above, we arrive at the distribution
of the (scaled) Wigner delay time $\tilde{\tau}_{W}$ for
the one-channel lead attached
to a site belonging to the $A-$ sublattice:
\begin{eqnarray}\label{main3l}
{\cal P}_l(\tilde{\tau}_{W})=\frac{1}{l!}
\frac{1}{\tilde{\tau}^{l+2}_{WA}}e^{-\frac{1}{\tilde{\tau}_{WA}}}
\end{eqnarray}
which generalizes the expression Eq.(\ref{main3}). In
Fig.~(\ref{fig-tau}) we present the distribution of the Wigner
delay times calculated numerically for various values of $l$
compared with this analytical result.

In fact, the time-delay distributions Eqs. (\ref{main3}) and
(\ref{main3l}) can be obtained in a more direct way. For this we
notice that the expression Eq.(\ref{smatrix})
for the unitary scattering matrix
$S_{A}(E)$ at real energies $E$ can be written in an equivalent form:
\begin{equation}\label{sae}
S_{A}(E)=\frac{\det{\left(E^2-i E P_{j_A}-\hat{H}
\hat{H}^{\dagger}\right)}}
{\det{\left(E^2+i E P_{j_A}-\hat{H}\hat{H}^{\dagger}\right)}}
\end{equation}
where we used the $N_A\times N_A$ projection matrix $P_{j_A}=
|j_A><j_A|$.
Calculating now the Wigner delay time as the energy derivative
of the logarithm of the above equation, and setting $E=0$ we arrive
after straightforward manipulations to a simple formula:
\begin{equation}\label{twa}
\tau_{W}=2<j_A|\left(\hat{H}\hat{H}^{\dagger}\right)^{-1}|j_A>=2\sum_{n=1}^{N_A}
\frac{|u_n(j_A)|^2}{\lambda^2_n}
\end{equation}
Comparison of this expression with Eq.(\ref{start}) makes it clear
that the Laplace transform $P(s)$ of the probability density for
$\tilde{\tau_{W}}=\tau_{W}/2N_A$ can be obtained via the same
manipulations as those leading to Eqs.(\ref{det}) and
(\ref{avres}). For the general case $0\le l=N_B-N_A\ll N_A,N_B$ we
find:
\[
P(s)=\left\langle
\frac{\det{\left(\hat{H}\hat{H}^{\dagger}\right)}}
{\det{\left(s/N^2_A+\hat{H}\hat{H}^{\dagger}\right)}}
\right\rangle_{H}=\frac{2}{l!} s^{\frac{l+1}{2}}
K_{l+1}(2\sqrt{s})
\]
which can be immediately inverted yielding the
distribution Eq.(\ref{main3l}).

It is easy to repeat the same manipulations for the lead attached
to a site of the larger sublattice $B$ and to find that
Eq.(\ref{twa}) is replaced formally with
\begin{equation}\label{twb}
\tau_{W}=2<j_B|\left(\hat{H}^{\dagger}\hat{H}\right)^{-1}|j_B>
\end{equation}
This expression, however, does not make any sense beyond the case
$l=0$ of equivalent sublattices, since the corresponding
$N_B\times N_B$ matrix $\hat{H}^{\dagger}\hat{H}$ is singular.
We conclude therefore that the Wigner delay time at zero energy
diverges for the sites belonging to the larger sublattice due to
presence of zero modes.

In summary, we presented analytical as well as numerical results
for the distribution of the local density of states for
zero-dimensional chaotic system at the point of chiral symmetry
$E=0$, and further utilized it for studying statistics of the
reflection coefficient and Wigner delay times. Apart from
investigating a gradual breakdown of the chiral symmetry with
increasing energy parameter $E$ in ergodic systems, an interesting
and challenging question is to study the LDOS fluctuations in
chiral systems of higher dimensions. In particular, in disordered
quantum wires the mean DOS is known to display a logarithmic
singularity\cite{Tit,AM}. It would be interesting to understand
how that singularity affects the LDOS distribution, and how it is
reflected in fluctuations of the related scattering
characteristics.

YVF is grateful  to LPTMS, Universite Paris-Sud
for kind hospitality and financial support during initial stage of
this work. Financial support by Brunel University VC grant (YVF) and  by a
grant from  the German-Israeli Foundation for Scientific Research and
Development (AO) is gratefully acknowledged.

\end{document}